\documentclass[preprint,aps,floats,showpacs,amssymb,tightenlines]{revtex4}
\usepackage{epsfig}
\usepackage{color}
\usepackage{amsfonts}
\usepackage{amsmath}

\newcommand{\beq}{\begin{equation}}
\newcommand{\eeq}{\end{equation}}
\newcommand{\be}{\begin{equation}}
\newcommand{\ee}{\end{equation}}
\newcommand{\beqs}{\begin{eqnarray}}
\newcommand{\eeqs}{\end{eqnarray}}
\newcommand{\bea}{\begin{eqnarray}}
\newcommand{\eea}{\end{eqnarray}}

\newcommand{\bml}{\begin{mathletters}}
\newcommand{\eml}{\end{mathletters}}

\newcommand{\beast}{\begin{eqnarray*}}
\newcommand{\eeast}{\end{eqnarray*}}

\newcommand{\vphi}{\varphi}

\begin{document}
\title{Charged-Rotating Black Holes in Higher-dimensional (A)DS-Gravity}
\renewcommand{\thefootnote}{\fnsymbol{footnote}}

\author{Y. Brihaye\footnote{brihaye@umh.ac.be} , T. Delsate\footnote{terence.delsate@umh.ac.be}}
\affiliation{D\'epartement de Physique Th\'eorique et Math\'ematiques ,
Universit\'e de Mons, Place du Parc, 7900 Mons, Belgique}  
\date{\today}
\setlength{\footnotesep}{1.5\footnotesep}

\begin{abstract}We present numerical evidences for the existence of rotating black hole solutions 
in d-dimensional Einstein-Maxwell theory with a cosmological constant and for $d$ odd.
The metric used possesses $(d+1)/2$ Killing vectors and the solutions have $(d-1)/2$ equal angular momenta. 
 A Schwarschild-type coordinate is used for the radial variable and both signs of the
cosmological constant are emphasized. Several properties of the solutions are studied, namely
their surface gravity, mass and angular momentum as functions of two parameters~:
the magnetic field and the angular velocity at the horizon.
The influence of the electromagnetic field on the domain of existence of the black holes
is studied are compared to the vacuum case where analytic solutions are available.
\end{abstract}

\maketitle
\section{Introduction}
Black holes constitute one of the most exciting predictions of four-dimensional general relativity (GR)
and the astrophysical detections of them constitute a great advance in physics. Black holes further played
a major role in establishing several mathematical aspects of GR and in  understanding  the physical 
interpretations  of its solutions \cite{hawking_penrose,israel,carter}.
 For numerous reasons, the interest for general relativity and of gravity in higher dimensions 
 has increased considerably in the last years (see e.g. \cite{peet},\cite{kknl},
 \cite{obers},\cite{charmousis}).  The question of the existence
 of black holes in higher dimensions therefore occurred naturally. One striking feature of the Einstein equations
 in more than four dimensions is that many uniqueness properties holding in four dimensions are lost.
For example, black holes presenting an horizon of topology $S_{d-2}$ can be constructed
explicitely in arbitrary dimensions  \cite{tangherlini};
their rotating generalisations \cite{mp}, the Myers-Perry (MP) solutions, are characterized by several
angular momenta. However several other types of solutions exist in higher dimensional gravity.
Black strings are one of them; they denote  string-like generalisations of 4-dimensional black holes  
in d-dimensional Einstein gravity; in particular, they are characterized 
by an event horizon of topology  $S_{d-3} \times S_1$ \cite{bs}.
The simplest black strings are independant on the extra coordinate, say $y$. Later on it was
realized that nonuniform solutions (depending on $y$) exist as well \cite{wiseman}; these
can further be done rotating \cite{kkr}.

Among possible solutions in higher dimensional gravity, we should also mention black ring \cite{ring}, with toroidal horizon topology and black saturn \cite{saturn}, where a black hole is surrounded by a black ring etc. These black object result from a balance between rotation along the torus and gravitational attraction and are known to be connected to the Myers Perry black hole in a limit where the rotation becomes high \cite{ring, saturn}. However, it has been found recently that black saturns can also exist without rotation in Einstein Maxwell theory \cite{satrad}. In this solution, it is the charge that balances gravity instead of rotation.

As a consequence, the classification of localized objects (black holes, black strings or other)
in higher dimensional gravity poses a serious mathematical problem.

Recently,  there has been  a lot of interest for d-dimensional GR with a cosmological constant $\Lambda$.
In the case of a negative cosmological constant, this interest  is 
motivated by the correspondance between the gravitating fields in an AdS space-time and 
 conformal field theory on the boundary of the AdS space-time \cite{Witten:1998qj,Maldacena:1997re}.
The propotype of higher-dimensional black holes with cosmological constant are
 known in an explicit form  \cite{Gibbons:2004js}. They generalize the solutions of \cite{tangherlini,mp}.
 Further explicit solutions are known in non minimal model e.g. \cite{ks,clp} or in the presence
 of a dilaton \cite{sheykhi,sheykhi1}.
As far as we know,  rotating black holes for the Maxwell field
 minimally coupled to Einstein gravity do not exist
in a closed form and one has to rely on perturbative or numerical methods to construct them 
respectively in asymptotically flat \cite{Kunz:2006eh,aliev0}, anti-DeSitter \cite{knlr,aliev}
and DeSitter \cite{bri_del} space-times.
The numerical construction reported in \cite{Kunz:2006eh,knlr} uses the isotropic coordinates, making the 
comparaison with the explicit solution of \cite{mp} (usually expressed in Schwarzchild coordinate)
not easy. In particular, the  deformation of the MP analytic solution due to the electromagnetic field 
can hardly be estimated as explained later in the text.
On the other hand, for $\Lambda > 0$, the occurence of a cosmological horizon leads to technical
difficulties which do not allow a direct comparaison between the domain of existence of the MP solutions
and those of charged black holes for both signs of the cosmological constant.

The purpose of this paper is to reexamine the equations of rotating black holes in d-dimensional 
Einstein gravity with a cosmological constant and with a minimal coupling to the electromagnetic field.  
We can then  study the effects both of the sign of $\Lambda$
and of the electromagnetic field on the MP solutions.  The charged rotating solutions exist for angular momenta higher than some critical momenta. When this  critical value is approached, the solution becomes extremal. However, we were able to address some extremal solution numerically by implementing a suitable set of boundary conditions. Extremal solution are interesting on their own, in particular, they are known to present symmetry enhancement \cite{enhanc} and play an important role in supergravity theory \cite{exsugra}.

We consider odd values $d$,  the ansatz for the metric is inspired from 
 \cite{mp,Kunz:2006eh}. It contains $(d+1)/2$ killing vectors, namely time-translation and $(d-1)/2$ related
 to rotations in suitable two-planes of space. Assuming all angular momenta in the two-planes
 to be equal and completing the ansatz by an appropriate form for the electromagnetic fields, we can
 transform the Einstein-Maxwell into a system of six differential equations.
 The remaining  arbitrariness in the coordinate system  is fixed by using  a Schwarzchild-like 
 radial variable. Charged, rotating black holes can then be constructed numerically
 by fixing three parameters by hand, namely the values of the horizon, the angular velocity 
 and  the magnetic field at the horizon. The challenging question is to determine
 the domain of existence of the solutions in this three-parameter space.

Although the equations are basically the same for both signs of the cosmological
constant, the pattern of the solutions can be considerably different. 


 The paper is organized as follow~: in Sect. 2 we present the model, the ansatz and the
 relevant one-dimensional reduced effective action (the equations are written in an Appendix).
 In Sect 3 we discuss the explicit solutions known in some limits, namely the non rotating case
 and the vacuum case. The boundary conditions are
 presented in Sect. 4,
   together with some physical quantities characterizing the solutions.
 The numerical results are described in Sect. 5 and illustrated by several figures. 
 In particular, the domain of existence of the black holes is emphasized for both 
 $\Lambda > 0$ and $\Lambda <0$ and compared to the vacuum case.  
 Finally, some conclusions are drawn in Sect. 6.

\section{General formalism}
\subsection{The Action}
We consider the Einstein-Maxwell equations  with a cosmological 
constant $\Lambda$ in a $d-$dimensional spacetime. The equations follow from the variations of the action 
\begin{eqnarray}
\label{action-grav}
I=\frac{1}{16 \pi G_d}\int_M~d^dx \sqrt{-g} (R - 2\Lambda- F_{\mu \nu}F^{\mu \nu})
-\frac{1}{8\pi G_d}\int_{\partial M} d^{d-1}x\sqrt{-h}K,
\end{eqnarray}
with respect to the metric and the electromagnetic field.
Here $G_d$ denotes the d-dimensional Newton constant and the units are chosen in such a way
that $G_d$ appears as an overal factor.
The last term  is the Gibbons-Hawking surface
term \cite{Gibbons:1976ue}. It
is required for the variational principle to be well-defined . The factor 
$K$ represents the trace
of the extrinsic curvature for the boundary $\partial\mathcal{M}$ and
$h$ is the induced
metric on the boundary.
It is standard to parametrize the cosmological constant by means of an (anti)-DeSitter radius $\ell$
according to  $\Lambda= \epsilon (d-2)(d-1)/(2\ell^2)$; $\epsilon=+1$ and $\epsilon=-1$  corresponds 
respectively to an asymptotic DeSitter and  
  Anti-DeSitter space-time.

\subsection{The ansatz}

To obtain rotating black hole solutions,
representing charged $U(1)$ generalizations of the MP
 solutions 
we consider space-times with odd dimensions: $d=2N+1$ and we parametrize  the metric  
 by means of an Ansatz
previously used for asymptotically flat solutions \cite{mp, Kunz:2006eh}
\begin{eqnarray}
 ds^2 = -b(r)dt^2 +  \frac{ dr^2}{f(r)} +
g(r)\sum_{i=1}^{N-1}
  \left(\prod_{j=0}^{i-1} \cos^2\theta_j \right) d\theta_i^2
  \nonumber \\
 + h(r) \sum_{k=1}^N \left( \prod_{l=0}^{k-1} \cos^2 \theta_l
  \right) \sin^2\theta_k \left( d\vphi_k - w(r)
  dt\right)^2
\nonumber
 \\
 +p(r) \left\{ \sum_{k=1}^N \left( \prod_{l=0}^{k-1} \cos^2
  \theta_l \right) \sin^2\theta_k  d\vphi_k^2 \right.
  -\left. \left[\sum_{k=1}^N \left( \prod_{l=0}^{k-1} \cos^2
  \theta_l \right) \sin^2\theta_k   d\vphi_k\right]^2 \right\} \ ,
  \label{metric}
\end{eqnarray}
In the above formula  $\theta_0 \equiv 0$ and  $\theta_N \equiv \pi/2$ are assumed;
the  non trivial angles have      $\theta_i \in [0,\pi/2]$
for $i=1,\dots , N-1$, while
 $\vphi_k \in [0,2\pi]$ for $k=1,\dots , N$. The corresponding space-time possesses
 $N$ commuting Killing-vectors
$\eta_k=\partial_{\varphi_k}$ supplementing the standard invariance of time translations $\partial_t$.
 The functions $b,f,h,g,w$ depend on the variable $r$,
the consistency of the ansatz imposes $p(r)=g(r)-h(r)$.

The most general Maxwell potential consistent with the symmetries
of the  line element (\ref{metric}) reads
\begin{eqnarray}
\label{Maxwell}
 A_\mu dx^\mu=V(r)dt+a_\varphi(r) 
\sum_{k=1}^N
\left( \prod_{l=0}^{k-1} \cos^2\theta_l \right) 
\sin^2\theta_k d\vphi_k
\end{eqnarray}
Where the electric and magnetic potentials $V(r)$ and  $a_\varphi(r)$ depend on $r$.


Substituting the ansatz (\ref{metric}),(\ref{Maxwell}) into the action (\ref{action-grav})
leads to an effective one-dimensional  reduced  Lagrangian   
\begin{eqnarray}
S_{red}\int dr ({\cal L}_{E}^{eff}+ {\cal L}_{M}^{eff})
\end{eqnarray}
where the Einstein part ${\cal L}_E^{eff}$ and the Maxwell part ${\cal L}_M^{eff}$ read respectively~:
\begin{eqnarray}
\label{Lg}
{\cal L}_E^{eff}=(d-3)g^{\frac{(d-7)}{2}}\sqrt{\frac{bh}{f}}((d-1)g-h)
+\frac{1}{2}\sqrt{fhb}g^{\frac{(d-3)}{2}}
(\frac{b'}{b}+(d-3)\frac{g'}{g})(\frac{h'}{h}+(d-3)\frac{g'}{g})
\\
\nonumber
-\frac{1}{4}(d-2)(d-3)\sqrt{fhb}g^{\frac{(d-7)}{2}}g'^2
+\frac{1}{2}g^{\frac{(d-3)}{2}}h\sqrt{\frac{fh}{b}}w'^2
-\epsilon \frac{(d-2)(d-1)}{\ell^2}g^{\frac{(d-3)}{2}}\sqrt{\frac{bh}{f}}
\end{eqnarray}
\begin{eqnarray}
\label{LM}
{\cal L}_M^{eff}=\frac{g^{\frac{(d-7)}{2}}}{\sqrt{bfh}}
\bigg(
2b\left(
2(d-3)a_\varphi^2h+fg^2a_\varphi'^2
\right)
-2fg^2h(w a_\varphi'+V')^2
\bigg).
\end{eqnarray}
The Einstein-Maxwell equations obtained from the ansatz (\ref{metric}),(\ref{Maxwell}) can be obtained in a standard way.
For the numerical construction of the solutions, the radial variable $r$
has to be specified  by fixing the "metric gauge";  
we find it  convenient to fix this arbitrariness   by choosing $g(r)=r^2$, 
defining a  radial coordinate $r$ of the Schwarzchild-type.
The corresponding equations are written explicitely in an Appendix.
Similarly to the asymptotically flat case  \cite{Kunz:2006eh}, the electric potential 
can be eliminated from the equations 
by making use of the first integral
(\ref{fiwbh}). The cosmological constant can be arbitrarily redefined by means of a 
rescaling of the radial variable $r$ and of the fields $h,V$ and $w$. For the numerical analysis,
we  use this arbitrariness to choose $r_h=1$ without loosing generality.



\section{Known solutions} 
\begin{figure}[!htb]
\centering
\leavevmode\epsfxsize=12.0cm
\epsfbox{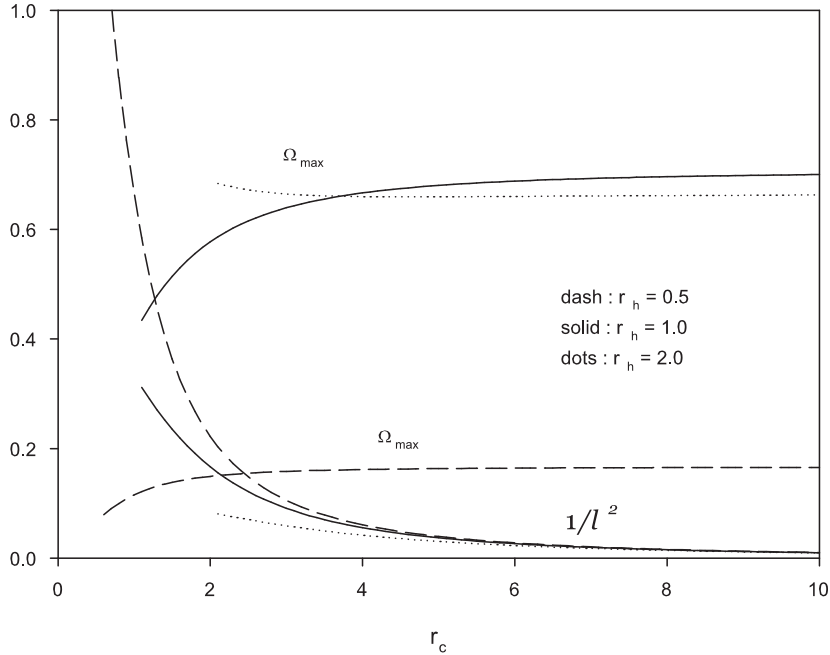}\\
\caption{\label{fig1} 
The value $\Omega_{max}$ and the parameter  $1/ \ell^2$
as functions of  $r_c$ for the MP solution and for $r_h=0.5,1.0,2.0$ }
\end{figure}

Solutions of the system of equations (\ref{ec2})-(\ref{ec7}) are known in two particular cases.
\\
(i) The de Sitter-Reissner-Nordstrom black holes \cite{amr,liu_sabra} are recovered in the limit $w(r)=a_\varphi(r)=0$~:
\begin{eqnarray}
f(r)&=&b(r)= 1 - \epsilon \frac{r^2}{\ell^2} - \frac{2 M}{r^{d-3}} + \frac{q^2}{2(d-2)(d-3) r^{2(d-3)}}  \ \ \ , \nonumber \\
 \ \ h(r)&=&g(r)=r^2 \ \ , \ \ 
V(r) = \frac{q}{(d-3)r^{d-3}}
\label{rn} 
\end{eqnarray}
where $M$ and $q$ are the mass and electric charge of the solution.
\\
(ii) The vacuum black holes discussed in \cite{Gibbons:2004js}
are recovered for a vanishing gauge fields ($V=a_{\varphi}=0$) and
\begin{eqnarray}
\label{vacuum}
 f(r)=1
-\epsilon \frac{r^2}{\ell^2}
-\frac{2M\Xi}{r^{d-3}}
+\frac{2Ma^2}{r^{d-1}},~ 
h(r)=r^2(1+\frac{2Ma^2}{r^{d-1}}),~ \nonumber  \\
w(r)=\frac{2Ma}{r^{d-3}h(r)},~~
g(r)=r^2,~~ b(r)=\frac{r^2f(r)}{h(r)},
\end{eqnarray}
where $M$ and $a$ are two constants related to the solution's mass and 
angular momentum and $\Xi=1+a^2/\ell^2$. 
These are generalisations of the Myers-Perry (MP) solutions \cite{mp}
to the case of non-vanishing cosmological constant. Here and in the following, 
 $\Omega \equiv w(r_h)$ will denote  the angular velocity at the event horizon.
 \\
%
%
Some  properties of the MP solutions  are useful for 
the discussion of charged rotating black holes in the case $\Lambda > 0$.
If we consider uncharged black holes solutions with fixed $r_h$ and $r_c$,
the explicit form provided by the Myers-Perry solutions reveals that rotating solutions
exist up to a maximal value of $\Omega$. For $d=5$, we find 
\be
    \Omega_{max}= \sqrt{\frac{2}{r_c^2+2 r_h^2}} \ 
    \frac{r_c r_h^3(r_c^2+r_h^2)}{r_c^2 r_h^4+r_c^2+ 2r_h^2}  \ \ , \ \ 
      \frac{1}{\ell^2}= \frac{1}{r_c^2+2r_h^2}
\ee
For $\Omega  \to \Omega_{max}$ the event horizon $r_h$ becomes a  double root of the function
$f$, i.e. the horizon becomes extremal. 
In fact for $\Omega > \Omega_{max}$, a third horizon occurs in the interval $[r_h,r_c]$ in the neighbourhood
of $r_h$. 
The values $\Omega_{max}$ and the corresponding value of the DeSitter radius $1/\ell^2$ are reported  
 as functions
of the cosmological horizon $r_c$ for three different values of $r_h$ on Fig.\ref{fig1}.

For fixed values of $r_h$, $r_c$, the value
of $\ell^2$ varies only a little for $\Omega \in [0, \Omega_{max}]$ ; e.g. for $r_h=1$,$r_c=3$ we have 
$1/10 \leq 1/ \ell^2 \leq  1/11$.
The mass $E$ and the angular momentum $J$ of the Myers-Perry solutions are given by \cite{gipepo}, \cite{kunduri}
\be
     E = \frac{V_{d-2}}{4 \pi G} M (\frac{d-2}{2} - \epsilon \frac{a^2}{2 \ell^2}) \ \ , \ \ 
     J = \frac{V_{d-2}}{4 \pi G} \frac{d-1}{2} M a 
\ee
where $V_{d-2}$ denotes the area of the $d-2$ dimensional sphere.
Throughout this paper, we will use the expression $E$ above to characterize the mass of the solutions
although a more elaborated formalism exists \cite{Balasubramanian:1999re,Brown:1993}. Once applied to
the MP solutions \cite{ghezelbash_mann}, it leads to an expression of the form
      ${\cal M}^{(d)} = E_c^{(d)} - \epsilon E$ where $E_c^{(d)}$ is interpreted as the Casimir energy
      of the dual Conformal Field Theory living on the $d-1$-dimensional boundary of the space-time \cite{Balasubramanian:1999re,Brown:1993}. The contribution $E_c^{(d)}$ appears only for odd values of $d$
      and depends on the space-time dimension and of the cosmological constant  \cite{klemm}; for $d=5$, one finds 
      $E_c^{(5)} = \frac{V_{d-2}}{4 \pi G} \frac{3 {\ell}^2}{16} $

\section{Boundary conditions and physical quantities}
\subsection{Constraints of regularity about the horizon}
We are interested in black hole solutions, with an horizon
located at $r=r_h$, i.e. with $b(r_h)=f(r_h)=0$.
Expanding the different functions  in the neighbourhood of the horizon 
\begin{eqnarray}
\label{expansion}
b(r)=b_1(r-r_h)+O(r-r_h)^2,
~~f(r)=f_1(r-r_h)+O(r-r_h)^2, \nonumber
\\
h(r)=h_0+h_1(r-r_h)+O(r-r_h)^2,
w(r)=w_h+w_1(r-r_h)+O(r-r_h)^2, \nonumber \\
 a_\varphi(r)=a_0+a_1(r-r_h)+O(r-r_h)^2,~~
V(r)=V_0+V_1(r-r_h) + O(r-r_h)^2
\end{eqnarray}
and inserting the expansions (\ref{expansion}) into the equations (\ref{ec2})-(\ref{ec7}) leads to terms proportional to $1/(r-r_h)$. 
For the solutions to be regular at the event horizon $r_h$ the coefficients of all these singular terms should vanish.
Since our numerical analysis will deal mainly with the case $d=5$, $\epsilon=1$, we write these 
rather cumbersome conditions in this case only.  
The equation for the function $h$ leads to the condition $\Gamma_1(r=r_h) = 0$ with
\beqs
  \Gamma_1(r) \equiv && 8 b' h^2 (12 a_{\varphi}^2 + 7 h) + 4 r 
b' h h' (12 a_{\varphi}^2 + 5 h) \nonumber \\
                     &-& 32 b' h^2 r^2 + 8 r^3 b' h (f' h - 4 h')         \nonumber \\
                     &+& 2 h r^4 ( \frac{24}{\ell^2} b' h^2 - f' 
                     (4 (a_{\varphi}')^2 h w^2 + 
8 a_{\varphi}' h w V' + b' h' + 5 h^2 (w')^2 + 4 h (V')^2  )   
                     )                     \nonumber \\
                     &+&  h' r^5 ( - \frac{24}{\ell^2} b' h^2 - f' 
                     (4 (a_{\varphi}')^2 h w^2 
+ 8 a_{\varphi}' h w V' - b' h' -  h^2 (w')^2 + 4 h (V')^2  )   
                     )   
\eeqs
where, for brevity, we dropped the dependence on $r$ in the various functions.  
The value $f'(r_h)$ appearing in the condition $\Gamma_1(r_h)=0$  
can be extracted from the equation for $f$, giving
\beq
      f'(r_h) = \frac{4 b_1 h_0}{r_h^3} 
      \frac{6 r_h^4/ \ell^2 + 8 r_h^2 - 12 a_0^2 - 5 h_0}
      {8 b_1 h_0 + r_h(4h_0(V_1+a_1 w_0)^2 - b_1 h_1 - h_0^2 w_1^2)}
\eeq
In the same way, the two Maxwell equations lead to a single condition
$\Gamma_2(r = r_h) 
= 0$ with
\beq
\Gamma_2 (r) \equiv 4 a_{\varphi} b' h + r^4 f'(h w' V' + a_{\varphi}' h w w' 
- a_{\varphi}' b')
\eeq
Similar conditions should be imposed at the cosmological horizon $r_c$ as well~:  
$\Gamma_1(r_c)=\Gamma_2(r_c)=0$.

\subsection{Physical quantities at the horizon}
The Hawking 
temperature and the event horizon area of these solutions can be obtained in a standard way, leading to 
\begin{eqnarray}
T_H=\frac{\sqrt{ f_1b_1}}{4\pi},~~
A_H=V_{d-2}r_h^{d-2}.
\end{eqnarray}
The mass and the angular velocity at the horizon are also usefull. They are defined by
means of the appropriate Komar integrals over the sphere $S^3_H$ at the event horizon. It leads to
\be
\label{komar}
   M_H = \frac{V_{d-2}}{8 \pi G_d} \sqrt{\frac{f h g^2}{b}} (b' - h w w')\vert_{r=r_h} \ \ , \ \ 
   J_H = \frac{V_{d-2}}{8 \pi G_d} 2\sqrt{\frac{f g^2 h^3}{b}} \  w' \vert_{r=r_h}
\ee
These quantities can be evaluated from the numerical solutions.
Similar quantities, say $M_C, J_C$ can be associated with the cosmological horizon. In fact, the masses and angular
velocities at the two horizons can be related to each other by formulas of the  Smarr-type.
Considering a Killing vector $K_{\mu}$,  integrating the two sides of the identity
\begin{equation}
              \nabla^{\mu} \nabla_{\mu} K_{\nu} = - R^{\mu} _{\ \nu} K_{\mu} = 
              - 8 \pi (T^{\mu}_{\ \nu} - \frac{1}{d-2} T \delta^{\mu}_{\nu}) K_{\mu}
\end{equation}
over the (truncated) hyper-volume $\Sigma$ covering the space-like region between the two horizons and
using both  Einstein equations and  Stokes theorem appropriately result into the identities we are interested in.
Choosing first $K = \partial_{\varphi}$ one gets
\be
   J_C - V_{d-2} \sqrt{\frac{fhg^2}{b}}(w a_{\varphi}+ V')\vert_{r=r_c} =  
   J_H - V_{d-2} \sqrt{\frac{fhg^2}{b}}(w a_{\varphi}+ V')\vert_{r=r_h} \ \ ,  
\ee
expressing the conservation of the total angular momentum.
The identity associated with the time-translation invariance ($K = \partial_t$)
is a bit more involved, it leads to
\be
M_C + \Phi_C Q_C - M_H - \Phi_H Q_H = \frac{1}{\pi \ell^2} V_{d-2} \int_{r_h}^{r_c} \sqrt{\frac{fhg^2}{b}} dr
\ee
where $Q_{H}$  (the electric charge  at the horizon \cite{maeda})
and $\Phi_H$ (the electrostatic potential)  are defined respectively by 
\be
          Q_{H} = \int_0^{\pi/2} d\theta (\sqrt{\frac{fhg^2}{b}})\vert_{r=r_h} \ \ \ , \ \ \ 
          \Phi_H = (V + \frac{d-1}{2} \Omega a_{\varphi})\vert_{r=r_h}
\ee
and similarly for $Q_R$, $\Phi_R$. Similarly to \cite{Kunz:2006eh}, the mass at the horizon can further
be expressed in terms of the surface gravity $\kappa$ and angular momentum according to
\be
   \frac{d-3}{d-2} M_H = \frac{1}{8 \pi G_d} \kappa A_H + \frac{d-1}{2} \Omega J_H
\ee
The various identities above were checked on our numerical solutions. 
\subsection{Asymptotics and global charges}
The asymptotic form of the solutions to the Einstein equations can be
constructed after some algebraic manipulations, leading to
\begin{eqnarray}
\label{asym}
b(r)=- \epsilon \frac{r^2}{\ell^2}+1+ \frac{\alpha}{r^{d-3}} +O(1/r^{2d-6}),
~~~
f(r)=- \epsilon \frac{r^2}{\ell^2}+1+\frac{\beta}{r^{d-3}} +O(1/r^{d-1}),
\\
\nonumber
h(r)=  r^2(1+  \epsilon \frac{\ell^2(\beta-\alpha)}{r^{d-1}} +O(1/r^{2d-4})),
~~~
w(r)=    \frac{\hat J}{r^{d-1}} +O(1/r^{2d-4 }). 
\end{eqnarray}
They depend on three arbitrary constants $\alpha,~\beta$ and $\hat J$.
The choice of the gauge metric $g=r^2$ in the ansatz (\ref{metric})
and the resulting functions $f,~b,~h$ have the advantage to present a 
direct connection with the closed form of the MP rotating solution.
The asymptotic expression of the gauge potential is similar to the asymptotically flat case
\begin{eqnarray}
 V(r)=- \frac{q}{r^{d-3}} +O(1/r^{2d-4 }),~~
 a_\varphi(r)= \frac{\hat \mu}{r^{d-3}} +O(1/r^{2d-4}) .
\end{eqnarray}

Although it is not clear if the thermodynamical properties are well defined in
the presence of the positive cosmological horizon, one may still define them in the standard way. 
In this section, we follow the lines of \cite{knlr} to present the global charges
characterizing the solutions asymptotically. 
ùùù\cite{Balasubramanian:1999re,Brown:1993}. 
 The mass-energy $E$ of the solutions and the angular momentum $J$ associated with an angular direction 
 are respectively given by
\begin{eqnarray}
\label{grav-charges}
 E=\frac{V_{d-2}}{16\pi G_d}(\beta-(d-1) \alpha),
~~J=\frac{V_{d-2}}{8\pi G_d}\hat J \ .
\end{eqnarray}
The above relations can be obtained namely by using a background subtraction approach
or the counterterm formalism \cite{Balasubramanian:1999re,Brown:1993}. Here do not take into account
the contribution of the Casimir energy of the solution \cite{klemm,ghezelbash_mann}.

The electric charge and the magnetic moment of the solutions are given by
\begin{eqnarray}
\label{gauge-charges}
 Q=\frac{(d-3)V_{d-2} }{4\pi G_d}q,~~~\mu=\frac{(d-3)V_{d-2}}{4\pi G_d}\hat \mu~.
\end{eqnarray}


\subsection{Boundary value problem}

\subsubsection{Case $\Lambda <0$}
In the case of asymtotically ADS black holes, the formulation of the boundary problem
is simple. Fixing $\Lambda <0$, the equations can be integrated in one step on the interval $r\in [r_h,\infty]$ 
by imposing  the boundary conditions 
\beq
        f = 0 \ , \ b = 0 \  , \ w = w_h \ , V = 0 \ , \ a_{\varphi}' = a'_h \ , \Gamma_1 = 0 \ , \ \Gamma_2 = 0
\  \ {\rm for} \ \ r = r_h \ ,
\eeq
\beq
  \ b = \frac{r^2}{\ell^2} + 1 + O(1/r^2)\ , \ h = r^2 \ ,  \ (r^2 a_{\phi})' = 0 ,
\  \ (r^4 w)'  = 0 \ {\rm for} \ \ r \to \infty  \ .
\eeq
The arbitrary additive constant of the electric potential has been used to set $V(r_h)=0$.
The parameters $w_h$, $a'_h$ have to be fixed by hand (alternatively, $a_{\varphi}' = a'_h$ can be replaced 
  by $a_{\varphi} = a_h$). The asymptotic decay of the electric potential $V$ and of the metric
function $f$ can then be used for  crosschecking  the numerical solutions. 
\\
Extremal black holes are characterized by a double zero of the functions $f,b$
at $r=r_h$. In the case of de Sitter-Reissner-Nordstrom black holes, analytical form
follows from (\ref{rn}), the parameters $M$, $q$ are then fixed as functions of $r_h$.
In the case of rotating solutions, the regularity conditions at $r=r_h$, leads to a set of rather 
involved conditions at $r=r_h$.

\subsubsection{Case $\Lambda >0$}
The positive cosmological constant leads to the occurence of
a $\Lambda$-depending cosmological horizon at $r=r_c$  with $r_h < r_c < \infty$ and where $f(r_c) = b(r_c)=0$. 
This creates several difficulties, namely~: 
(i) the point $r_c$ constitutes a  singular point of the equations,
(i) the point $r_c$ is not known a priori as a function of $\Lambda$,  
(iii) imposing regularity at both horizons $r_h$ and $r_c$ needs twelfe boundary conditions, 
this number exceeds  the eleven boundary conditions allowed by orders of the six equations
(for instance, one of first order and five of second order).

In order to overcome these difficulties, we supplement the system (\ref{ec2}),(\ref{ec7})
by the trivial equation $d \ell^2 /dx = 0$.
The new system then needs twelve boundary
conditions to be well posed. The strategy consists in  solving the equations in two steps. First we choose by hand
the event horizon $r_h$ and the cosmological horizon $r_c$ and integrate the equations on the interval
$[r_h,r_c]$ by imposing the regularity conditions at the two horizons. It turns out that this needs
exactly twelve boundary conditions, namely~:  
\beq
        f = 0 \ , \ b = 0 \ , \ b' = 1 \ , \ w = w_h \ , V = 0 \ , \ a_{\varphi}' = a'_h \ , \Gamma_1 = 0 \ , \ \Gamma_2 = 0
\  \ {\rm for} \ \ r = r_h
\eeq
\beq
        f = 0 \ , \ b = 0 \ ,   \Gamma_1 = 0 \ , \ \Gamma_2 = 0
\  \ {\rm for} \ \ r = r_c
\eeq
The functions $\Gamma_{1,2}$ are defined above.
Here, we choose the arbitrary rescaling of the time variable $t$ in order to impose $b'(r_h)=1$.  
The function $b(r)$ is therefore not normalized according to (\ref{asym}), the appropriate
normalisation, say $b \to b/{\cal N}_b^2$, can be obtained only after the second step. 
The constants $w_h$ and $a'_h$ are  arbitrary
and control the total angular momentum and the (electric and magnetic) charges of the black hole.
The numerical value of $\ell^2$ is determined by the first step, together with the values of
all the different fields at $r=r_c$. 

The second step consists in integrating the equations on $[r_c,\infty]$. 
The value of $\ell^2$ is known from the first step and the numerical values of the
fields at $r=r_c$ can 
be used as a  set of Cauchy data to solve the equations from $r=r_c$ to  $r \geq r_c$. 

This method presents several features of inconvenience. 
Namely: (i)
  a systematic analysis of the
solution for a fixed value of the cosmological constant cannot be performed. Fortunately, the numerical value
of $\ell^2$ depends only a little on $w(r_h)$ and $a_{\varphi}(r_h)$ once $r_h,r_c$ are fixed. 
(ii) The appropriate normalisation of $b$,$w$,$V$ leading to an asymptotically DeSitter metric
is only known a posteriori. 
$$
    b \rightarrow \frac{b}{{\cal N}_b^2} \ \ , \ \ w \rightarrow \frac{w}{{\cal N}_b} \ \ , \ \ 
    V \rightarrow \frac{V}{{\cal N}_b}  
$$
In particular it turns out impossible to study solutions with a fixed value of the 
angular velocity at the horizon $\Omega$ (since $\Omega = w(r_h)/{\cal N}_b$)
and varying the horizon. In the numerical analysis, we put the emphasis on the families of solutions
obtained when varying $\Omega$ for fixed values of the horizons $r_h,r_c$ and of the parameter $a_h'$.

\section{Numerical results}
The system of equations (\ref{ec2})-(\ref{ec7}) does not admit, to our knowledge, explicit solutions for
generic values of $w_h$ and $a'_h$. We therefore rely on a numerical method to
construct solutions. We solved the equations in the case $d=5$ and we hope that this case
catches the qualitative properties of the pattern of the solutions available for $d>5$. The numerical solver
Colsys \cite{colsys} was used to obtain the results. 
\\
\subsection{Charged solutions for $\Lambda < 0$}
Charged 5-dimensional rotating black hole solutions have been presented in details in \cite{knlr}
where the coordinate freedom is fixed by chosing the isotropic coordinate, say $y$.
The solutions are constructed as function of $y$ and their pattern is discussed mainly for fixed horizon at $y=y_h$.
With this definition of the radial coordinate, the pattern of the solutions available for 
fixed $y_h$ present several branches. For instance, solutions with different masses but same $y_h$ and $\Omega$
can be found.
Unfortunately, the solutions can hardly be compared with the the explicit form (\ref{vacuum})
of the Myers-Perry solutions which hold with the Schwarzschild  (or the Boyer-Linquist) coordinate.
For this reason, and for the purpose of comparison of the solutions between the Anti-DeSitter and  DeSitter cases,
we present some solutions with $\Lambda <0$ in the Schwarzschild coordinate $g=r^2$ and determine the pattern of
solutions available with $r_h$ fixed. For all this section we set $r_h=1$ and $\ell^2 = 0.1$.
\begin{figure}[!htb]
\centering
\leavevmode\epsfxsize=12.0cm
\epsfbox{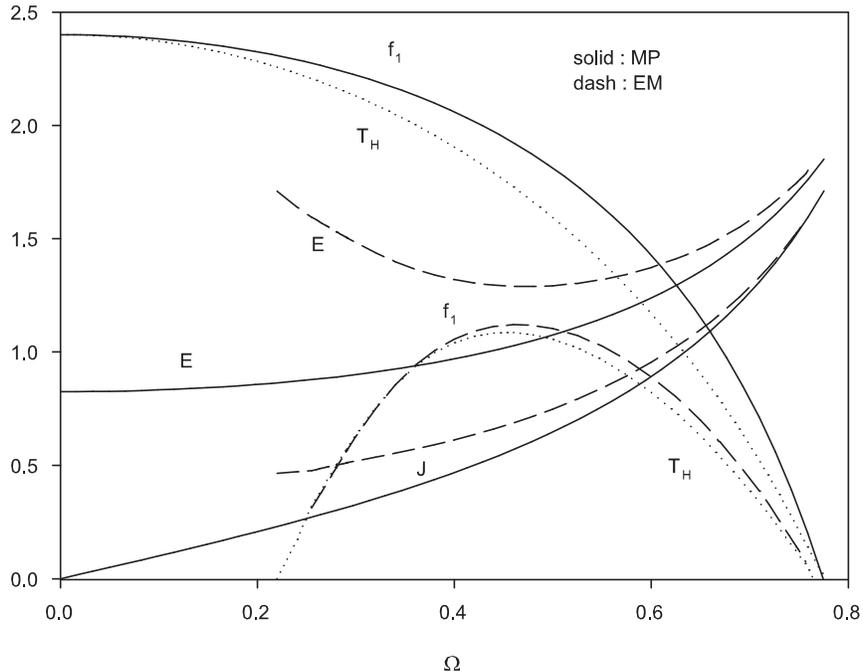}\\
\caption{\label{solu_ads} 
The Mass $E$, the angular momentum $J$ and the value $f_1$ 
 as functions of $\Omega$  for the Myers-Perry solution (solid) and the charged solution with $A'=0.5$ (dashed)
 for $r_h=1$ and $\ell^2=0.1$.
 The Hawking temperature $T_H$ is represented by the dotted lines}
\end{figure}
\begin{figure}[!htb]
\centering
\leavevmode\epsfxsize=12.0cm
\epsfbox{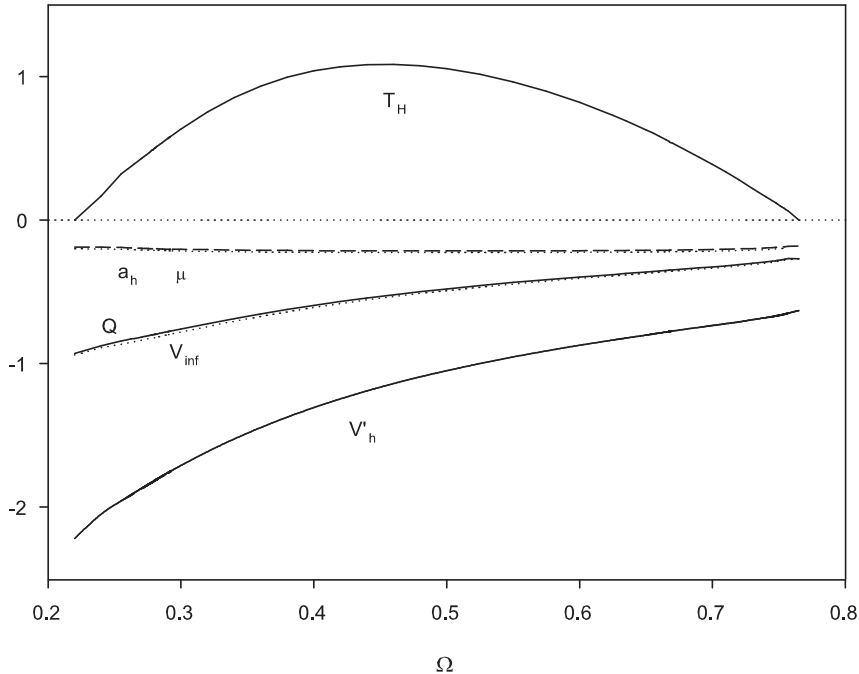}\\
\caption{\label{ads_1_c} 
Parameters of the electromagnetic field of charged ADS-BH 
as function of $\Omega$ for $r_h=1$, $a'_h=0.5$.
 }
 \end{figure}
Several physical parameters characterizing the charged rotating black holes  with $r_h=1$, $a'_h=0.5$ 
are represented on Fig. \ref{solu_ads} by the dashed lines.
The corresponding parameters
associated with the  MP solutions are  represented on this figure by the solid lines.
As already pointed out, vacuum (or MP) solutions exist for positive values of $\Omega$ up to a 
maximal value $\Omega_{max}$, with the choice of the parameters of the figure, we have $\Omega_{max} \sim 0.78$.
Our numerical integration of the full equations reveals that charged black holes exist only for a finite
interval of the horizon angular velocity~: $\Omega \in [\Omega_a, \Omega_b]$, with 
$\Omega_a \geq 0$ and $\Omega_b \leq \Omega_{max}$. The values  $\Omega_{a,b}$
depend on the strength of the electromagnetic fields. In the case of Fig. \ref{solu_ads} 
we find $\Omega_{a} \sim 0.22$, $\Omega_{b}\sim 0.765$.
In the limits $\Omega \to \Omega_{a,b}$,  
extremal solutions are approached, i.e. corresponding to $f_1=b_1=0$. As a consequence, the Hawking temperature 
$T_H$ also tends to zero in these extremal limits. It is represented by the dotted lines on Fig. \ref{solu_ads}.  
 \begin{figure}[!htb]
\centering
\leavevmode\epsfxsize=12.0cm
\epsfbox{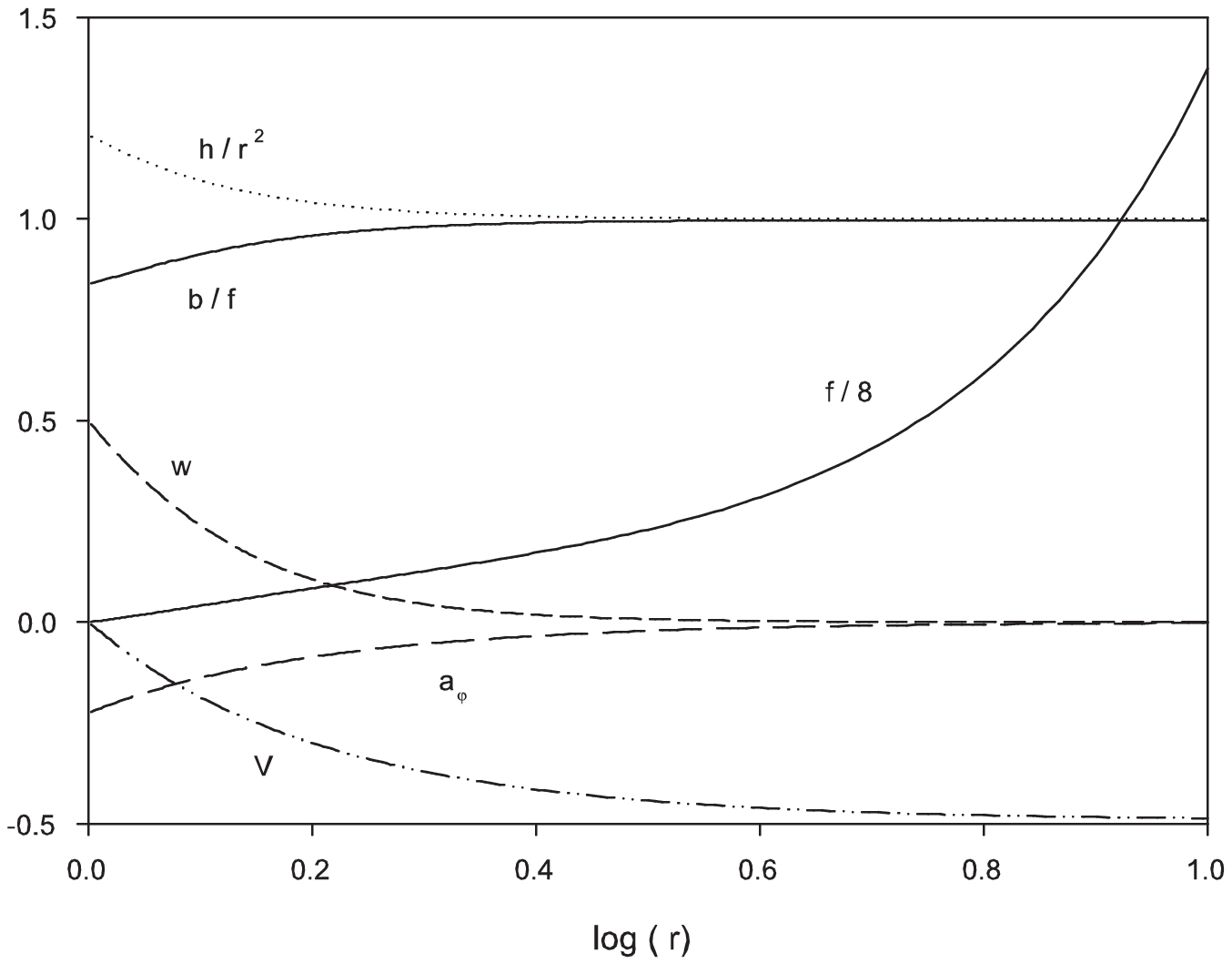}\\
\caption{\label{ads_2} 
The profiles of a typical ADS rotating black hole with $r_h=1$,$\Omega=0.5$ and $a'_h=0.5$
 }
 \end{figure}
Several parameters characterizing the electromagnetic fields of the solutions have also been computed  numerically.
The electric charge $Q$ and the magnetic moment $\mu$, vary slowly as functions of the horizon velocity 
$\Omega$, as shown by Fig. \ref{ads_1_c}.  
Note that that the numerical values
of $Q$ and $V(\infty)$ are very close to each other and similarly for $\mu$ and $A_h$.
Profiles of a typical ADS rotating black hole is presented on Fig. \ref{ads_2}. 
We managed to construct rotating extremal black holes numerically for small values of $\Omega$.
The profile of such a solution is presented in Fig. \ref{extre_bh5} corresponding to $\Omega=0.09$.
The numerical integration is delicate; a systematic construction if extremal solutions is not
aimed in this paper but would require an appropriate reparametrisation of the metric. 
 \begin{figure}[!htb]
\centering
\leavevmode\epsfxsize=12.0cm
\epsfbox{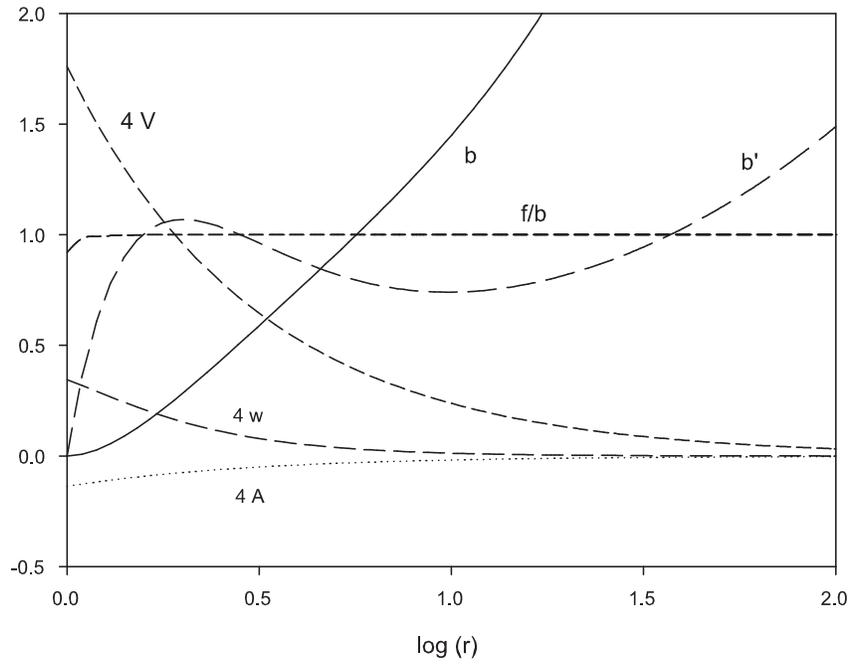}\\
\caption{\label{extre_bh5} 
The profiles of an  extremal ADS rotating black hole with $r_h=1$,$\Omega=0.09$ and $\Lambda = - 0.1$
 }
 \end{figure}
\subsection{Remarks on the Isotropic coordinate}
We would like to stress the main differences between the patterns of the solutions
once constructed   by using the isotropic coordinate and the Schwarzschild coordinate. 
\begin{figure}[!htb]
\centering
\leavevmode\epsfxsize=12.0cm
\epsfbox{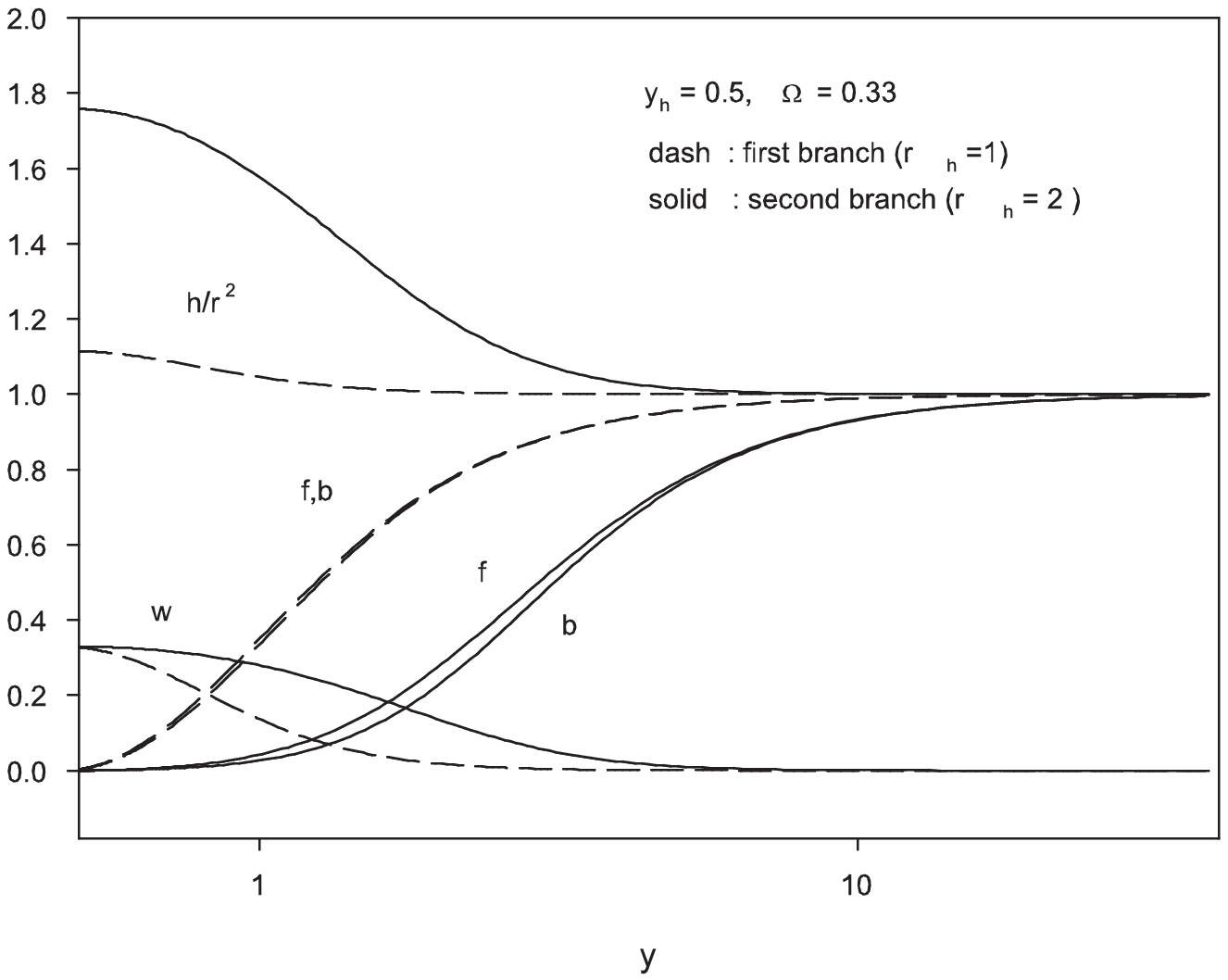}\\
\caption{\label{isotropic} 
The profiles of two black hole solutions corresponding to $y_h=0.6$, $\Omega=0.33$ }
\end{figure}
\begin{figure}[!htb]
\centering
\leavevmode\epsfxsize=12.0cm
\epsfbox{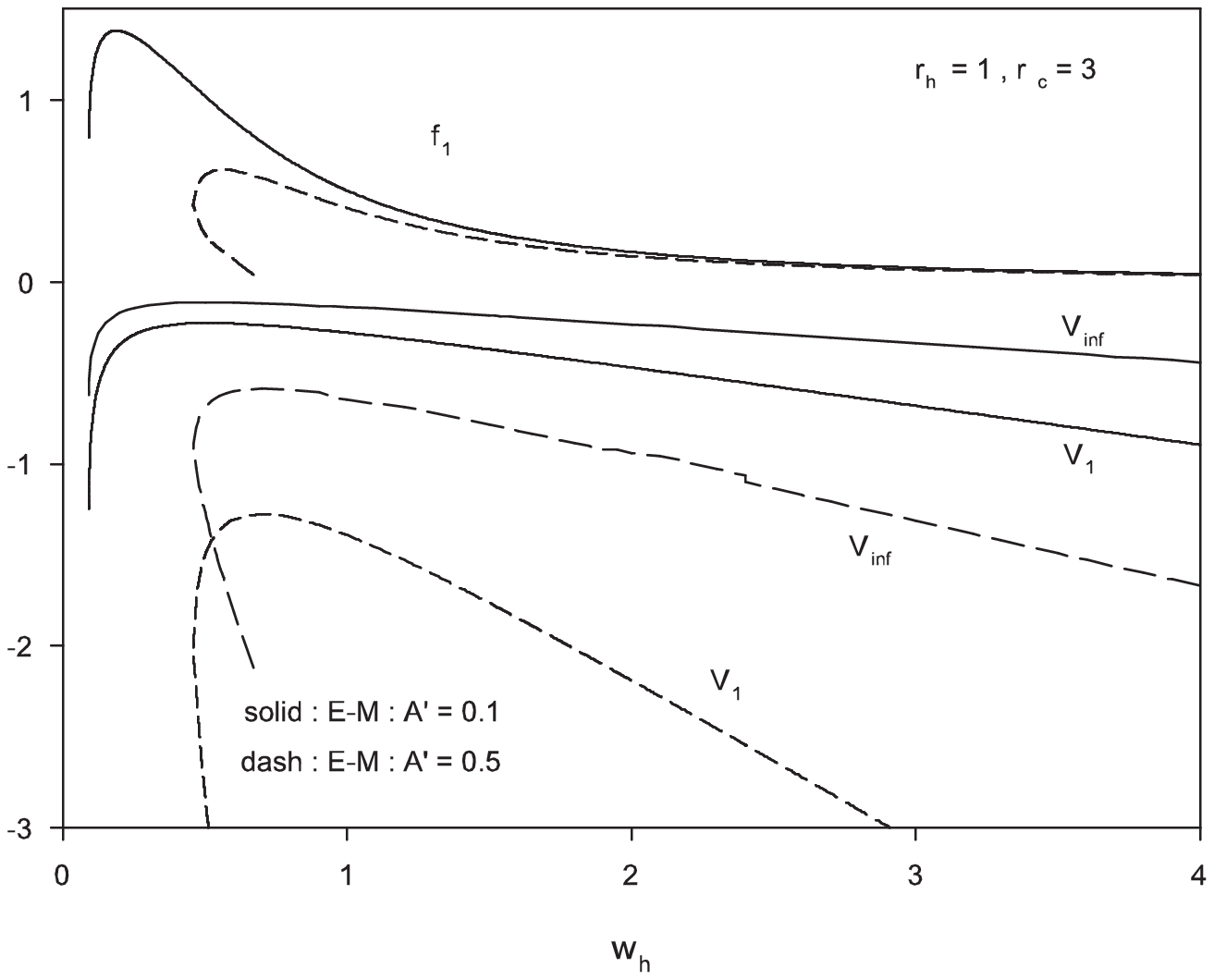}\\
\caption{\label{fig2c} 
The values $f_1$, $V_1$ and $V_{inf}\equiv V(\infty)$ are reported as functions of $w_h$ 
for $a_h'=0,0.1$ and $a_h'=0.5$
 }
 \end{figure}
We make these remarks in the vacuum, asymptotically flat  case~: $ \Lambda = 0$.
In the present paper, we use the Schwarschild coordinate $r$, denoting the angular velocity at the horizon
as $\Omega \equiv w(r_h)$ 
fixing the horizon $r_h$, we have solutions with a regular horizon
for $\Omega \leq \frac{1}{r_h \sqrt{2}}$. All physical quantities,  the mass in particular
are monotonic functions of $\Omega$. In contrast, the isotropic coordinate $y$ is used
e.g. in \cite{knlr}, we have
\be
     y = \exp{K(r)} \ \ , \ \ K'(r) = \frac{1}{r\sqrt{f}}
\ee
While families of solutions (MP) take a rather simple form in the Schwarzschild coordinate
it is not the case with the isotropic coordinate.
Considering the family of solutions with fixed horizon $y_h$ in the isotropic coodinate,
leads to a rather involved pattern in the $r_h$, $\Omega$ plane. Namely it can be that two 
solutions with equal $y_h$ and $\Omega$ can exist but with different masses.
This is illustrated on Fig. \ref{isotropic} where 
two solutions corresponding to $r_h \approx 1$ and $r_h \approx 2$ and $\Omega = 0.33$ have
both $y_h=0.5$ once expressed in the isotropic coordinate.
\begin{figure}[!htb]
\centering
\leavevmode\epsfxsize=12.0cm
\epsfbox{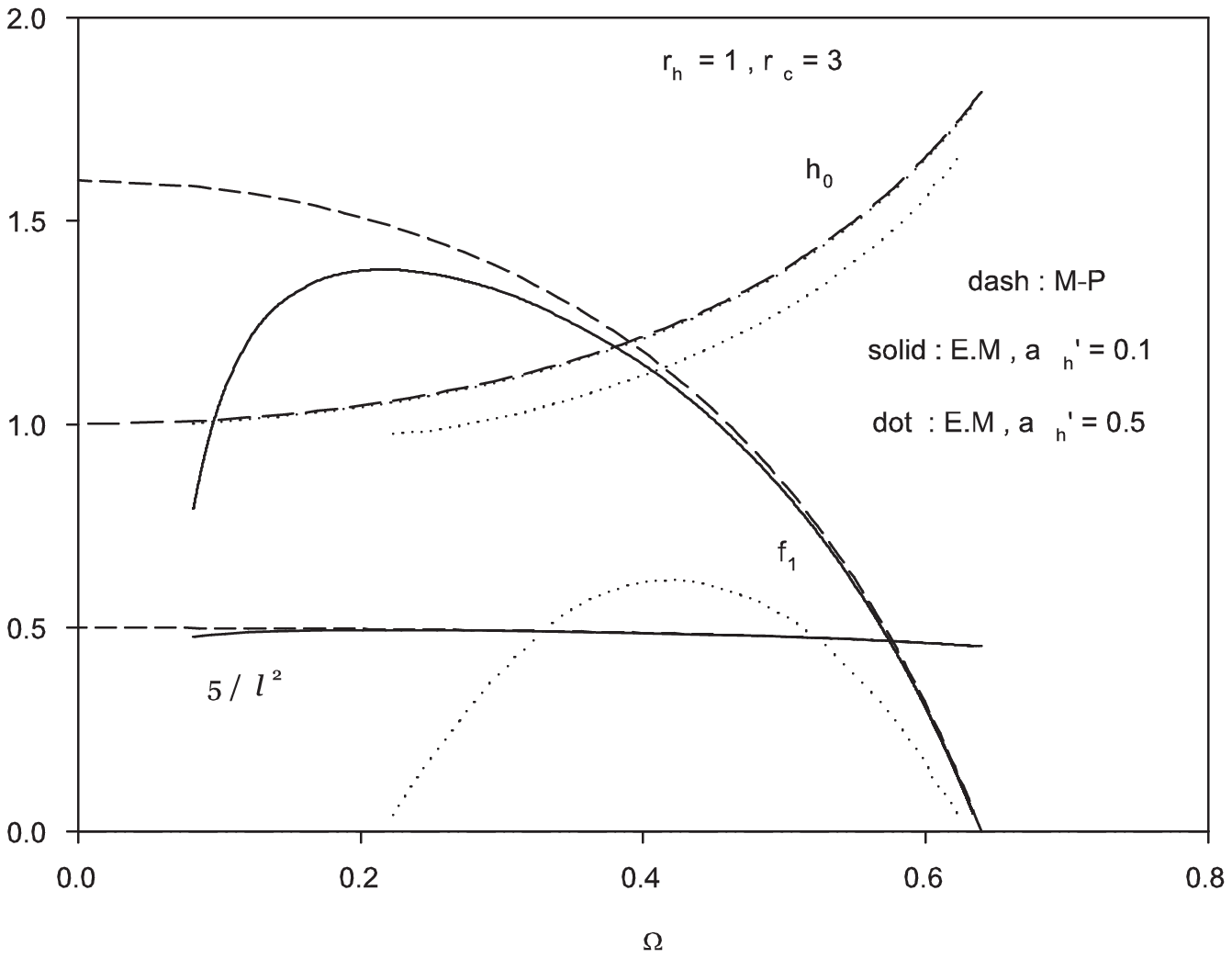}\\
\caption{\label{fig2} 
The value $f_1$ and $g_0$ are reported as functions of $\Omega$ 
for $a_h'=0,0.1$ and $a_h'=0.5$
 }
\end{figure}
\begin{figure}[!htb]
\centering
\leavevmode\epsfxsize=12.0cm
\epsfbox{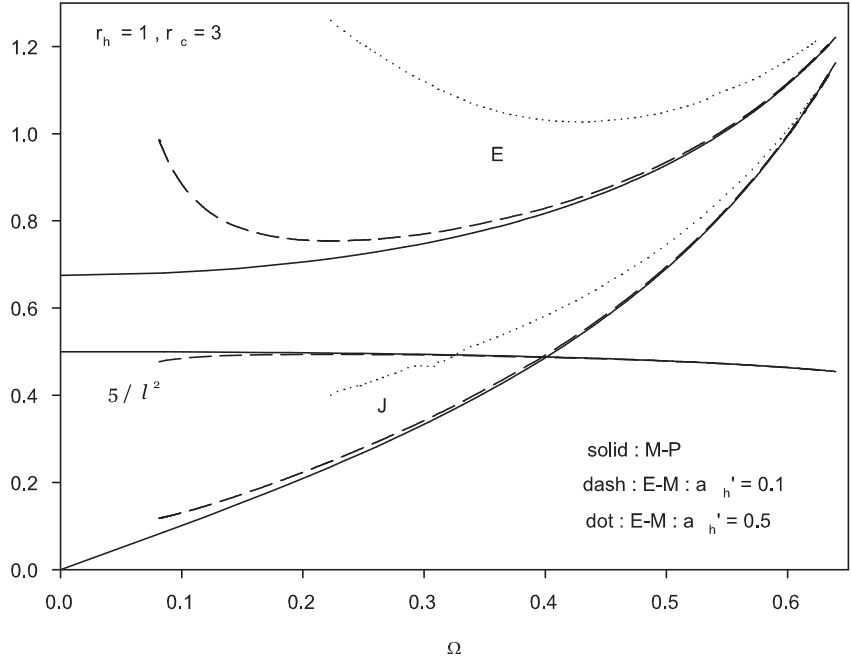}\\
\caption{\label{fig2b} 
The Energy of the solutions and the angular momentum $J$ are reported as functions of $\Omega$ 
for $a_h'=0 ; 0.1$ and $a_h'=0.5$
 }
\end{figure}
\begin{figure}[!htb]
\centering
\leavevmode\epsfxsize=12.0cm
\epsfbox{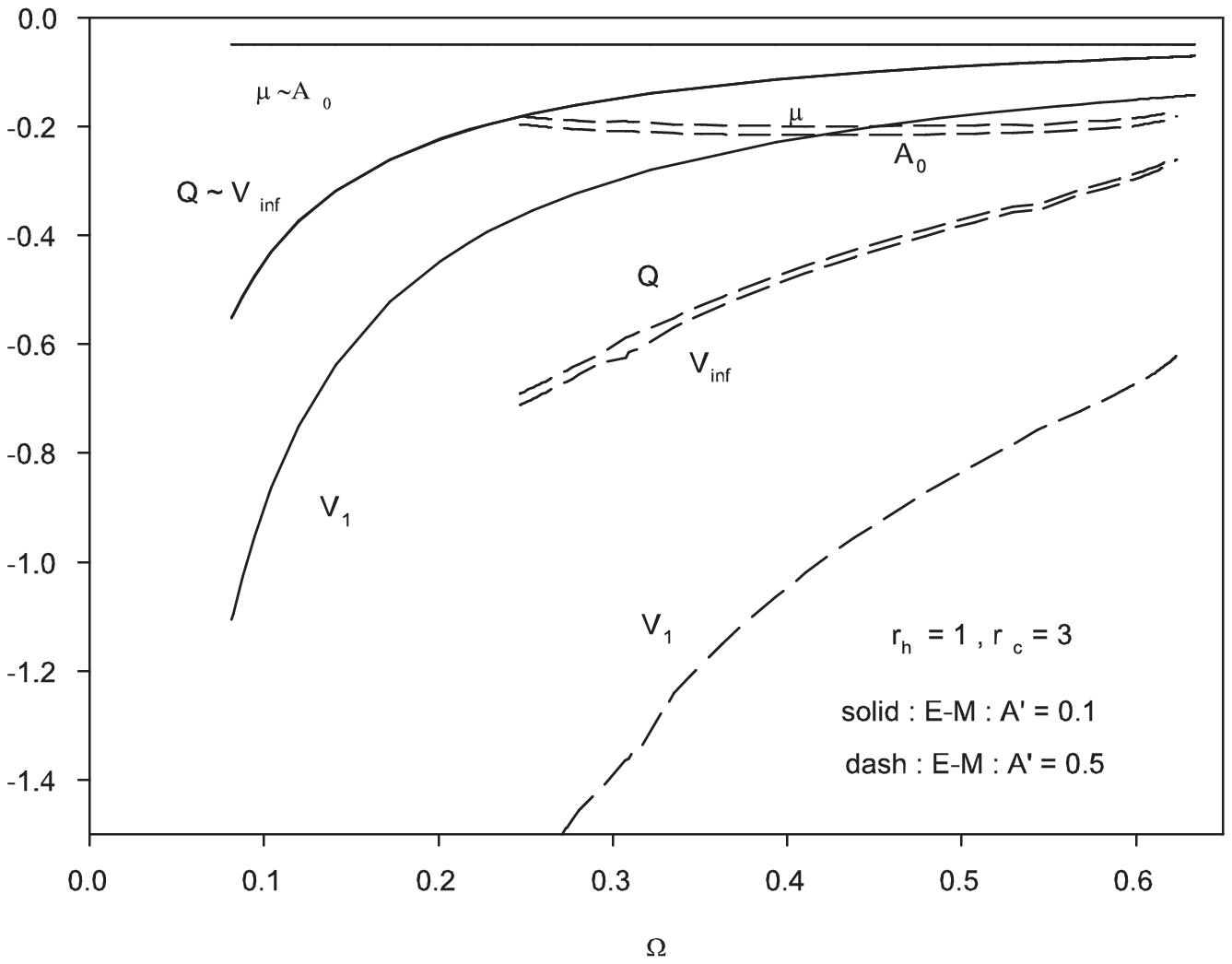}\\
\caption{\label{fig2d} 
Electromagnetic parameters  as functions of $\Omega$ 
for $a_h' 0.1$ and $a_h'=0.5$
 }
\end{figure}

\subsection{Charged solutions for $\Lambda > 0$}
For $\Lambda > 0$, the presence of a Maxwell field  changes the pattern of solutions corresponding 
to the vacuum (or MP) solutions. Along with the case $\Lambda < 0$,
our numerical analysis strongly suggests that charged-rotating black holes exist only for a finite
interval of the horizon angular velocity $\Omega$, i.e. for $\Omega \in [\Omega_a, \Omega_b]$  
with $0 < \Omega_a < \Omega_b < \Omega_{max}$. Here $\Omega_{max}$ denotes the maximal horizon
velocity for the MP solution and  the values $\Omega_{a,b}$
depend of the strength of the electromagnetic field. 
\begin{figure}
\centering
\leavevmode\epsfxsize=12.0cm
\epsfbox{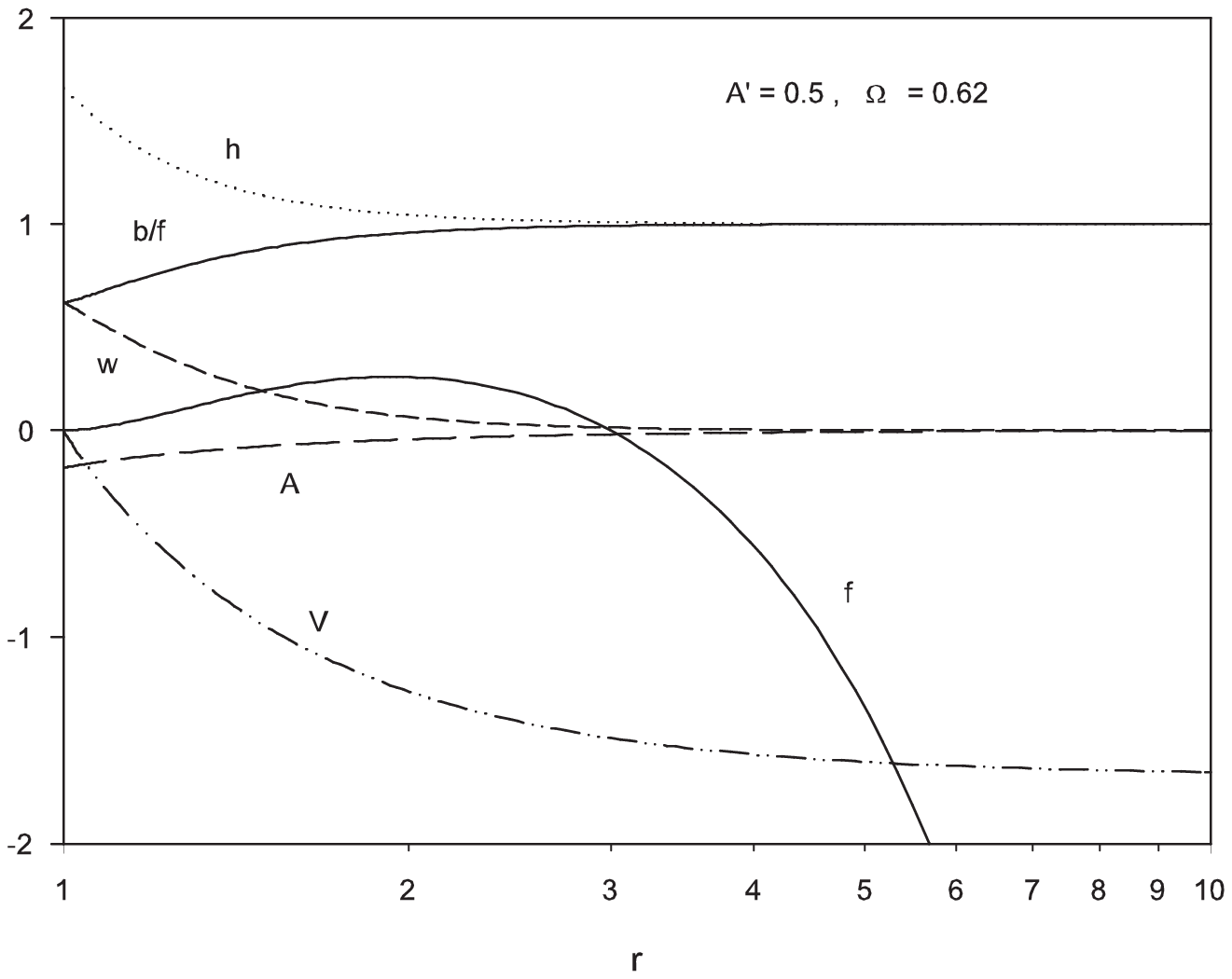}\\
\caption{\label{fig4} 
The profile of the metric and Maxwell function for $r_c=3,r_h=1$
for $a_h'=0.5$ and $\Omega=0.62$
 }
\end{figure}
\begin{figure}
\centering
\leavevmode\epsfxsize=12.0cm
\epsfbox{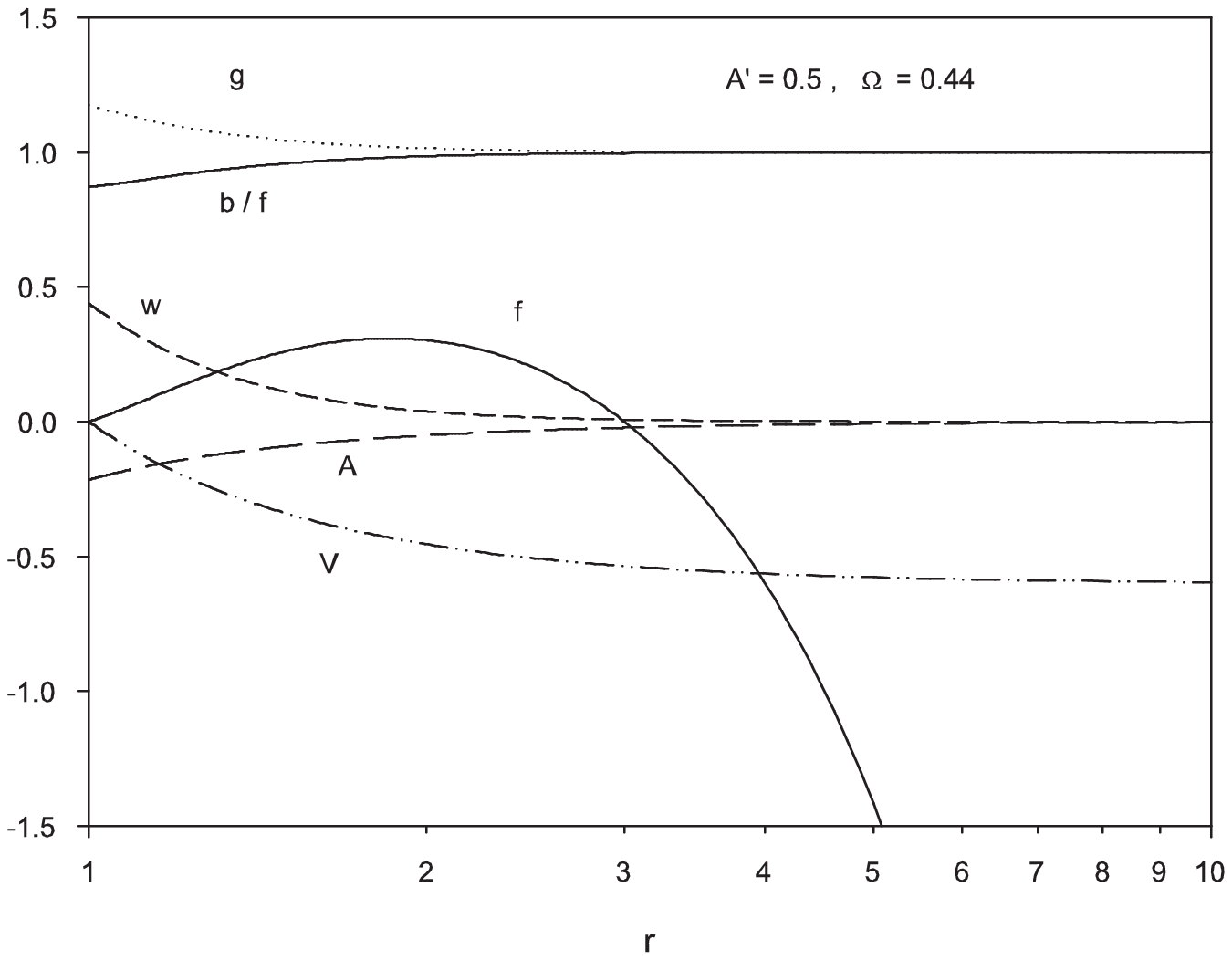}\\
\caption{\label{fig5} 
The profile of the metric and Maxwell function for $r_c=3,r_h=1$
for $a_h'=0.5$ and $\Omega=0.44$
 }
\end{figure}
It must be stressed that the numerical integration of the equations become
extremely tedious in the region of the parameters which allow to approach
the lower limit $\Omega \sim \Omega_a$.
The reason of this difficulty comes from the fact that, when the shooting parameter $w_h$  decreases,
the branch of solutions stops as some minimal value, say $w_{h,min}$. Then, a second branch of solution
emerge for $w_h \geq w_{h,min}$. The two branches coincide it the limit $w_h \to w_{h,min}$. Following
the second branch shows that it terminate into a critical value of $w_{h,c}$ and $f_1 \to 0$ for $w_h \to w_{h,c}$.
This is illustrated on Fig. \ref{fig2c} where a few parameters that are substantially different on the two branches
are represented as functions of $w_h$.
In all figures of this section, the values $r_h=1$,$r_c=3$ are used corresponding to $\ell^2 \sim 10$.
In fact the second branch is represented for $a'_h = 0.5$ (where we find $w_{h,min} \sim 0.459$ and  $w_{h,c}\sim 0.67$). 
A  second branch exist also for the other lines presented on the figure (i.e. corresponding to $a_h'=0.1$)
 but it turned out even more difficult to construct it with a good accuracy than for the case $a'_h=0.5$
 and we refrain to representing it. 
On the other side on the graphic, solutions seem to exist for arbitrarily large values of the parameter $w_h$.

So, once  described in terms of $w_h$, the pattern of solution looks complicated. 
However, it is  more relevant to  plot the different  quantities as functions of the physically meaningfull 
 parameter $\Omega$ representing the angular velocity at the horizon. 
This is obtained by renormalizing the fields $b,w,V$ appropriately so that the metric become asymptocally DeSitter. 
  
Several parameters  characterizing  the metric functions at the horizon  are reported on Fig. \ref{fig2}.
as function of $\Omega$.
The results summarized on this graphic suggest that the solutions  exist for $\Omega \in [\Omega_a, \Omega_b]$, 
and that in the two limits $\Omega \to \Omega_{a,b}$
an extremal black hole is  approached at the event horizon $r_h$. 
Physical parameters like the energy $E$ and the angular momentum $J$, which are related to the asymptotic charges,
 are reported on Fig.  \ref{fig2b}. The two figures  \ref{fig2}, \ref{fig2b} further show that,
for $a_h'$ fixed, the parameters of the charged black hole deviate more substantially
from the corresponding Myers-Perry solutions in the region of  small values of $\Omega$. 
We note that, for non-vanishing electromagnetic field there is only one solution for given  $\Omega$, $r_h$. 
This contrasts with
the description in term the isotropic coodinate where in general
two solutions can be associated to one value of $\Omega$ (see Fig.1 of \cite{knlr}).
Finally several parameters related to the Maxwell fields $V$ and $a_{\varphi}$ are
reported of Fig.\ref{fig2d}. We see that the values of $\mu$,$A_0$ which hardly differ for small $a_h'$ are
lifted when the amplitude of the electromagnetic field increases (and similarly for $Q$, $V_{inf}$). 

Profiles of two typical solutions having $a'_h=0.5$ are represented on Figs. \ref{fig4},\ref{fig5}
repectively for $\Omega = 0.62$, i.e. close to the limit $\Omega_b \sim 0.64$ and for an intermediate value
$\Omega = 0.44$

\section{Conclusions}
In this paper, numerical aguments are given for the existence of
five-dimensional charged and rotating black holes with a positive cosmological constant.
Two main features of these solutions are that (i) they are characterized by at least two horizons,
an event horizon $r_h$ and a cosmological horizon $r_c$; for fixed $r_h$,$r_c$, we find a families
of black holes characterized by the parameters $\Omega$ and $a'_h$ determining respectively 
the angular velocity and the magnetic fields at the
event horizon. We presented several arguments indicating that
black hole solutions exist only for $\Omega$ taking values in a finite
interval, say $[\Omega_a, \Omega_b]$. 
The limiting values depend on the strength of the
electromagnetic field (in our case this is controlled by 
parameter $a_h'$)  and determine the domain of existence of the black holes in the $\Omega,a'_h$ plane
(with $r_h,r_c$ fixed). 
The solutions end up into extremal black holes when the values $\Omega_{a,b}$ are approached. 
In the limit $a_h' \to 0$, the MP black holes are recovered, existing for $\Omega \in [0,\Omega_{max}]$
with $\Omega_{max}$ known analytically.
This feature is common to both signs of the cosmological constant once solution are constructed with
a Schwarschild -like coordinate. 
Smarr-type formulas have been obtained which relate some physical quantitities associated with the event
horizon and the cosmological horizon in the case of a positive cosmological constant.

\section{Appendix: The equations}
In this Appendix, we present the equations associated to the Einstein-Maxwell Lagrangian.
First, the equations determining the metric~: 
\begin{eqnarray}
\label{ec2}
f'
+\frac{f}{(d-2)}
\bigg(-\frac{rh}{2b}w'^2
+\frac{2r}{ b}V'^2
+\frac{4rw}{b}a_\varphi 'V'
+\frac{h'}{h}(1-\frac{rb'}{2b})
-2r(\frac{1}{h}-\frac{w^2}{b})a_\varphi'^2
+\frac{b'}{b}
\\
\nonumber
+\frac{(d-1)(d-4)}{r}
\bigg)
+\frac{1}{(d-2)r^3}
((3d-5)h+4(d+1)a_\varphi^2-(d-1)^2r^2)+\frac{(d-1)r}{\ell^2}
=0,
\end{eqnarray}
\begin{eqnarray}
\label{ec1}
b''+
 \frac{1}{d-2}
\bigg(
 4(5-2d)w a_\varphi'V'
 +\frac{(d-3)}{2h}b'h'
 -\frac{2(d-3)b}{2rh}h'
 -2(2d-5)V'^2
\\
\nonumber
 +\frac{1}{2}(3-2d)hw'^2
-2(\frac{b}{h}+(2d-5)w^2)a_\varphi'^2
+(d-2)\left(\frac{b'f'}{2f}-\frac{b'^2}{2b}\right)
+\frac{(d-3)^2}{r}b'
\\
\nonumber
-\frac{(d-3)b}{r^4f}(12 a_\varphi^2+h)
+\frac{(d-3)b}{r^2}\left(\frac{d-1}{f}+4-d\right)
+\frac{(d-1)(d-2)b}{\ell^2f}
\bigg)=0,
\end{eqnarray}
\begin{eqnarray}
\label{ec3}
h''
+\frac{1}{(d-2)}
\bigg(
\frac{(2d-5)h^2}{2b}w'^2
+\frac{2h}{b}V'^2
+\frac{4hw }{b}a_\varphi'V'
-\frac{(d-2)h'}{2}(\frac{h'}{h}-\frac{f'}{f})
\\
\nonumber
+\frac{(d-3)}{2b}b'h'
+\frac{(d-3)^2}{r}h'
+2(\frac{hw^2}{b}+2d-5)a_\varphi'^2
-\frac{(d-3)h}{rb}b'
-\frac{(d-3)(2d-3)h^2}{r^4f}
\\
\nonumber
-\frac{12(d-3)a_\varphi^2h}{r^4f}
-\frac{(d-3)(d-4)h}{r^2}
+\frac{(d-1)h}{f}(\frac{d-2}{\ell^2}-\frac{d-3}{r^2})
\bigg)=0,
\end{eqnarray}
\begin{eqnarray}
\label{ec4}
w''
-\frac{4w }{h}a_\varphi'^2
-\frac{4a_\varphi'V'}{h}
+\frac{(d-3)w'}{r}
+\frac{1}{2}\left(-\frac{b'}{b}+\frac{f'}{f}+\frac{3h'}{h}\right)w'=0,
\end{eqnarray}
%
%
%
%
%
%
 and, for the Maxwell fields
%
%
\begin{eqnarray}
\label{ec6}
V''
-\frac{w}{b}b'a_\varphi'
+\frac{w}{h}a_\varphi'h'
+\frac{1}{2}(\frac{2(d-3)}{r}-\frac{b'}{b}+\frac{f'}{f}+\frac{h'}{h})V' \nonumber \\
+(1+\frac{hw^2}{b})a_\varphi'w'
+\frac{hw}{b}V'w'
+\frac{2(d-3)a_\varphi hw}{r^4f}=0,
\end{eqnarray}
\begin{eqnarray}
\label{ec7}
a_\varphi''
+\frac{1}{2}(\frac{2(d-3)}{r}+\frac{b'}{b}+\frac{f'}{f}-\frac{h'}{h})a_\varphi'
-\frac{h}{b}(wa_\varphi'+V')w'
-\frac{2(d-3)a_\varphi h}{r^4f}=0.
\end{eqnarray}
It can easily be seen that the  equations of motion have the
first integral
\begin{eqnarray}
\label{fiwbh}
 g^{\frac{(d-3)}{2} }\sqrt{\frac{fh}{b}}(wa_\varphi'+V')=(d-3)q.
\end{eqnarray} 
\\
\\
\noindent {\bf Acknowledgments} 
We gratefully acknowledge the Belgian F.N.R.S. for financial support and Eugen Radu
for numerous discussions. 

\end{document}